\documentstyle[floats,aps]{revtex}
 \newcommand{\drp}[2]{\frac{\partial #1}{\partial #2}}
  \newcommand{\vv}[1]{{\bf \hat {#1}}}
\input epsf
\input rotate
\input psfig.sty

\begin{document}
\draft
\title{Vesicle propulsion in haptotaxis : a local model}

\author{Isabelle Cantat, Chaouqi Misbah and Yukio Saito}
\address{Laboratoire de Spectrom\'etrie Physique, Universit\'e Joseph Fourier
 (CNRS),
Grenoble I, B.P. 87, Saint-Martin d'H\`eres, 38402 Cedex, France
}
\address{Department of Physics, Keio University, 
 3-14-1 Hiyoshi, Kohoku-ku, Yokohama 223-8522, Japan
 }
\author{\parbox{397pt}{\vglue 0.3cm 
\small
We study theoretically vesicle locomotion  due to haptotaxis. 
    Haptotaxis is referred to motion induced by an adhesion gradient
    on a substrate.  The problem is solved
    within a local approximation where a Rayleigh-type dissipation
    is adopted. The dynamical model is akin to the Rousse model
    for polymers. A powerful gauge-field invariant formulation
is used to solve a dynamical model which includes a kind
of dissipation due to bond breaking/restoring with the substrate.
For a stationary situation where the vesicle acquires
a constant drift velocity, we formulate the propulsion problem
in terms of a nonlinear eigenvalue (the a priori unknown drift  velocity)
one of Barenblat-Zeldovitch type. A counting argument  shows
that the velocity belongs to a discrete set. For a relatively tense
vesicle, we provide an analytical expression for the drift velocity
as a function of relevant parameters. We find good agreement
with the full numerical solution. Despite the oversimplification of the model
it allows the identification of a relevant quantity, namely the 
adhesion length, which turns out to be crucial also in the  nonlocal
model in the presence of hydrodynamics, a situation on which
we have recently 
reported [I. Cantat, and C. Misbah, Phys. Rev. Lett. {\bf 83}, 235 (1999)]
and which constitutes the subject of a forthcoming extensive study.
}}
\author{\vskip 0.2 cm PACS numbers 87.22.Bt, 87.45 -k, 47.55 -Dz}

\maketitle

\section{Introduction}
 Phospholipidic vesicles constitute a simple model of cytoplasmic
 membranes of real cells. A simple model due to Helfrich\cite{Helfrich73}
 based on curvature energy has  accounted for a variety of equilibrium
 shapes. The model is based 
 on a minimal energy principle\cite{Lipowsky}. Some of the shapes (the so-called discocytes) bear 
 strong resemblance with that of an erythrocyte (the red blood cell).
 Additionally,
 analysis of flickering (temporal small fluctuations
 around a given shape) of an erythrocyte by Brochard and
 Lennon\cite{Brochard75} has been quite successfully
described by the Helfrich model including hydrodynamics dissipation.
 The vesicle  model has  seemed then as a natural
 candidate, at
 least in a first stage,
 for dealing with  more complex entities such
 as those encountered in the realm of biology.
 In that context, however, most of the features are of
 nonequilibrium dissipative nature. Very recently 
 several 
 theoretical\cite{Prost96,Kraus96,Durand97,Cantat99,Cantat99a,Seifert99,Cantat99b} 
 and experimental\cite{Nardi99} investigations have
 been directed along that line. 

We are interested here  in vesicle migration, a question
on which we have given recently a brief account\cite{Durand97,Cantat99}.
 Despite the very complex biochemical behaviour of a cell, cells may
  also exhibit behaviours where simple physical concepts may be evoked. 
  It is
 well documented that, for example, the migration of the pronephric duct cells in salamanders
  is regulated by haptotaxis. Haptotaxis is a terminology that is used
  to express the  following fact: when adhesive molecules
  are present in increasing amounts along an extracellular
   matrix (or simply on a substrate in {\it vitro}), a cell
   that was constantly making and breaking adhesion with such
    a molecule would migrate from a region of low
     concentration to an area where that adhesive molecule
     was more highly concentrated\cite{Carter67,Curtis69}.
     There are also evidences that cell migration during embryo development may be guided by
      an adhesion gradient. In other words cell migration
      is here guided by a purely external physical
      factor, while the internal structure (the cytoskeletton)
      is quite unaffected on the time  scale of interest.
      This feature drastically differs from that of a cell
      belonging to the immune system where the cytoskeletton plays
      a decisive role\cite{cytoskeletton}. Despite
      the fact that the cytoskeletton in pure haptotaxis does not
      undergo a structural change as is the case during cell
      crawling, 
      the problem remains very much involved
      since the cell cytoskeletton dissipation 
      may come to the fore as well as an intricate
      bond breaking and restoring with the substrate. We shall
      consider here a pure vesicle moving in haptotaxis. Our belief
      is  that advancement in this field can be achieved only by
      the progressive refinement of concepts.

      We consider a vesicle moving along the substrate thanks
      to an adhesion gradient. As the vesicle moves, it generates
      hydrodynamics flow both inside and outside. 
      Hydrodynamics induces nonlocal interactions leading
	    to an effective coupling of two
		  distinct regions on the vesicle.
		  In addition,
      the two
      monolayers that form the phospholipidic
      membrane might slide one relative to the other.
      Finally during motion the vesicle forms new bonds ahead and destroys
      others behind, and this process of bond breaking and restoring
      may be so slow that it may dominate dynamics (see later).

      This paper should be regarded as using a very simplistic
      view in the hope of introducing the concepts of migration
      and to exhibit in a transparent
      fashion the way the problem is addressed. We shall keep
      the description as simple as possible. That is to say: (i) we
      ignore nonlocality due to hydrodynamics --incorporation
      of hydrodynamics was briefly discussed in \cite{Cantat99} and
      will be the subject of a forthcoming paper--, (ii) we confine
      ourselves to a 2D geometry. 
      Some kind of dissipation due to bond breaking and restoring
      is introduced  in our model. The adoption of a local
      model (no hydrodynamics) allows one to quite easily  obtain analytical
      results and thus to extract some
      key ingredients about migration --especially the role
      of the adhesion area (length in 2D)-- which 
       turns out to be crucial also when hydrodynamics is
       included. 

       The scheme of this paper is organized as follows. In Section II
       we write down the equations of motion and comment them. In section
       III we present a forward time integration and present the main
       result. Section IV  presents the solution
       of the stationary system in a form of a nonlinear eigenvalue
       problem, where the drift velocity is the eigenvalue. In section V
       we give an analytical solution. A conclusion
       and a discussion is presented in section VI.

\section{Equation of motion}

\subsection{Parameterization}
We consider an adhering vesicle, deposited on a flat substratum which is
 oriented 
by its normal vector ${\bf \hat y}$ (Fig.1).
The x-axis is along 
the wall and represents the direction of vesicle motion occurring by convention 
from left to right.  
As stated before we confine ourselves to two dimensions.
That is to say, the vesicle morphology is invariant in the z-direction, similar 
to a tubular vesicle. 


The interaction between
the vesicle and  the substrate is taken into account by introducing
an adhesion potential. The range of the potential  in realistic
situations (typically several $nm$)  is small in comparison 
to the vesicle size (several $\mu m$), so that it is
justified in practice to consider a contact potential, unless
otherwise indicated (see later).
The energy interaction is then zero if $y > 0$ and is non vanishing only close
to contact (if $y=0$). 
At the junction point between the free part of the vesicle (whose
length is denoted as $L^*$) and the adhered part (with length
$L_{adh}$), the potential undergoes an abrupt change.
The  contact between the  membrane and substrate occurs
at two well-defined points $x_1$ at the left and $x_2$ at the right.
These  parameters are related to the total length of the curve $L$ by 
$L= L_{adh}+L^*= (x_2-x_1)+L^*$.
We use an intrinsic representation of the vesicle contour by
introducing $\psi(s)$,
the angle between the outward normal and the y-axis,  and $s$
the  arc length,
as shown on Fig. 1.
We only need to consider the function $\psi(s,t)$  from $s=0$ to  $s=L^*$ 
corresponding 
to the contact points $x_1$ and $x_2$,  respectively. Because
of the contact potential character  the adhesion length is completely fixed
if the  two contact points $x_1$ and $x_2$ are known.
Thus the vesicle shape and its dynamical properties (like the 
propulsion velocity) are known if $x_1$, $x_2$ and the 
function $\psi(s,t)$ are determined. The demand  that
the parametrization of the vesicle be compatible with the 
adhesion on the substrate it fulfilled by the 
two geometrical constraints :
(i) the distance between both contact points, $x_2-x_1$,  must
coincide with the adhesion length $L_{adh}=x _2-x_1$, (ii) 
their vertical coordinates $y_1$ and $y_2$ must have
the same value, 
 $y_2 -y_1=0$. These two constraints can be expressed in terms
of $\psi(s,t)$. For that purpose we use  the relations
\begin{equation}
\drp{x}{s}=\cos \psi \; , \;\;\;\;\;    \drp{y}{s}=-\sin \psi \; ,
\label{dxds}
\end{equation}
which allow us to write the two constraints in the 
following form :
 \begin{eqnarray}
\int_0^{L^*} \drp{x}{s} \, ds &=& \int_0^{L^*} \cos \psi(s)\, ds = x_2-x_1 = 
L_{adh} \; ,  \label{cos} \\
\int_0^{L^*} \drp{y}{s}\,  ds &=& \int_0^{L*} - \sin \psi(s) \, ds = y_2-y_1 = 0 
\; .
\label{sin}
  \end{eqnarray}
These are the geometrical constraints. In order to describe vesicle
dynamics, we need  a dynamical equation for the evolution
of  $\psi(s,t)$. A movement of the vesicle (due for example
to an adhesion gradient) is limited
by dissipation (such as hydrodynamics etc...). The vesicle
reacts to any deviation from equilibrium by its internal forces
(bending, possible stretching --or resistance to stretching--). 
Let us first discuss these forces.
\subsection{Energy and forces}

All the relevant membrane properties are summarized in the following energy, 
expressed in 2D, with the 
dimension of an energy per unit length : 
\begin{equation}
E=\int_{\cal C} \kappa  \frac{(c-c_s)^2}{2} ds-\int_{x_1}^{x_2}w(x)\,dx
+\int_{\cal C} \zeta(s) ds
+p\, S \; .
\label{Fad}
\end{equation}
The first term is the well known Helfrich curvature energy,
 with the rigidity $\kappa$, the curvature $c=\partial \psi/ \partial s$
and the spontaneous curvature $c_s$ \cite{Helfrich73}. 
Because of  the 2D-specific 
conservation law $\int c \,ds=2\pi$, any curve displacement leaves unchanged
 the energy terms associated to the 
 spontaneous curvature. We can thus omit
the term associated with $c_s$.
The second term expresses the adhesion energy and is only integrated on the 
adhered part of the curve. 
As we are concerned with an inhomogeneous substratum, the contact potential 
depends on the variable
 $x$ and is denoted by 
$-w(x)$ (with $w>0$, meaning that 
adhesion is favorable).
Finally the  last two terms ensure the  length and surface conservation, 
respectively. 
The membrane is a two dimensional {\it incompressible} fluid. 
The phospholipid exchanges 
with the solvent is virtually absent, and    
the area per molecule on the vesicle remains constant.
This  
leads to the local length conservation (in the 2D language). 
The variable $\zeta$ is  a local Lagrange multiplier which enforces
the arc length
$ds$  to a constant value. 
The enclosed surface $S$ conservation is a consequence of the membrane 
impermeability and fluid incompressibility. 
It is ensured  by    the global Lagrange multiplier 
$p$.
The interpretation of $\zeta$ and $p$ as a tension and a pressure will be 
discussed latter.
 
The functional
derivative of the  energy (\ref{Fad}) provides
us with the various forces acting on the membrane. 
  
\vspace{0.2cm} 
\noindent {\bf (i) Curvature forces} \\
 Under the assumption that the membrane is completely flat on the adhered part, 
we reduce
 the integration domain of the first energetic term in eq.(\ref{Fad}) to 
$[0,L^*]$.
 Making use of the relations ${\bf t}= {\bf r}'$ and ${\bf n}=- (1/c) {\bf r}''$ 
 (the prime designates derivative with respect to  $s$) 
 which are the membrane tangential and normal unit vectors, we obtain for 
 the curvature energy $E_c$
 \begin{equation}
 E_c={\kappa \over 2}\int_0^{L^*} \left ( \drp{^2{\bf r}}{s^2} \right )^2 ds \; 
.
 \end{equation}
 
 When taking the functional derivative of $E_c$ with respect
 to the position, care must be taken. 
 Indeed the arc length element  must also undergo a variation.
A convenient formulation avoiding  confusion rests on the introduction
of a  general
 parametrization $a \in [0,1]$,  related to $s$
by the metric $g=(ds/da)^2$, and is time-independent.  
The energy expression becomes\cite{Cantat99a,Cantatthes}
\begin{equation}
 E_c={\kappa \over 2} \int_0^1   \left[ \left( \drp{^2 {\bf r}}{a^2} \right )^2 
 -{1 \over g} \left( \drp{^2 {\bf r}}{a^2}  \drp{{\bf r}}{a}\right )^2 \right ]  
g^{-3/2} da \; . 
\label{Fdea} 
\end{equation} 
The functional derivative, though straightforward, may
be too lengthy if one does not take
care in regrouping adequately various
terms as explained in \cite{Cantat99a}.
The result can be written in a simple form:
\begin{equation}
{\bf f}_c= -{\delta E_c \over \delta {\bf r}(s)}=  \kappa \left (\drp{^2 c}{s^2} 
+ {c^3 \over 2 }\right )
 \, {\bf n} \; .
\label{fc} 
\end{equation}

The curvature force is, as expected, free  of any tangential contribution. 
 The first term in eq.(\ref{fc}), involving the second derivative of the
 curvature,
  tends to keep  curvature repartition as homogeneous 
as possible. It is also present in 3D under the more complicated form of the 
Laplace-Beltrami operator. 
The second term proportional to  $c^3$ is in contrast 
  2D-specific. It  tends to increase
the size of any  convex shape. Note that  in 3D the 
curvature energy is scale invariant, which implies a 
vanishing curvature force of this type  on a sphere. 
The difference between 2D and 3D can be explained in the following way. 
Let us consider  a finite cylinder of length $H$ and radius $R \ll H$,
which constitutes  a good approximation 
for an infinitely long cylinder. 
In order to make the cylinder "closer" 
to a sphere, which is the corresponding 3D equilibrium shape,
the curvature force would tend to increase the radius and 
decrease the length so as to bring the cylinder shape as close
as possible to a sphere.
 This  gives an  intuitive picture of the
$c^3$ term in 2D. \\

In the discussion above we did omit the boundary contribution
when taking the functional derivative. Since this point is a bit subtle
we have postponed it to the end of this section.

\vspace{0.2cm}
\noindent {\bf (ii) Length and surface constraints} \\
On the free part of the curve, the force  which is associated
to the first Lagrange multiplier $\zeta$ is obtained
upon functional derivation of 
$E_l=\int \zeta ds$. The result is
\begin{equation}
{\bf f}_l=-\delta E_l/\delta {\bf r}(s)= -c \, \zeta(s)\,{\bf n} +{d \zeta \over 
ds} \, {\bf t} \; .
\label{fl}
\end{equation}
The normal component is easily identified as a Laplace pressure, whereas 
the tangential one looks like a Marangoni force (which is encountered
when surface tension is inhomogeneous). 
However, $\zeta$ is not exactly  similar to a surface tension as
for an interface
between two fluids. 
The  "tension" $\zeta$ is not an intrinsic property of the membrane. It 
adapts itself to the other forces in order to maintain the local length
fixed. 
In other words,
the problem is implicitly written in a thermodynamical ensemble with fixed 
length. This differs from 
the usual problem  for fluid or solid surfaces
where the surface tension is fixed instead. 
 Thus $\zeta$ is a variable that  must be determined self consistently 
as a Lagrange multiplier, by use of the constraint equation (see 
appendix in Ref.\cite{Csahok99a}) :
\begin{equation}
0=\drp{(ds)}{t}=\left ( \drp{v_t}{s}+c v_n \right ) ds \; .
\label{divv}
\end{equation}
This relation (\ref{divv}) simply expresses the condition of vanishing velocity 
divergence on the curved contour
of the vesicle, 
which is precisely  the incompressibility condition in the 2D fluid
constituting the membrane (written here in one dimension). A more
intuitive way of viewing expression (\ref{divv}) is  presented on
Fig.2. 
The Marangoni term is the only tangential term among all membrane forces 
(see eqs. \ref{fc},\ref{fl},\ref{fs}).  It is seen from (\ref{fl})
that the Lagrange multiplier must be uniform at equilibrium.
For sake of simplicity and in order to get more insight 
into analytical understanding, a uniform 
value will be assigned to $\zeta$, even out-of-equilibrium.  A discussion
of this point will be presented in section VI.     
This assumption implies some consequences on dynamics (and especially
on the tangential velocity) which will be presented in 
section \ref{dynamics}.

\vspace{0.2cm}
Finally we have to consider the force associated to $E_s=p S$ :
\begin{equation}
{\bf f}_s=-\delta E_s/\delta {\bf r}(s)= -p\,{\bf n} \; .
\label{fs}
\end{equation}
The Lagrange multiplier $p$ depends only on  time; it enforces
a constant 
area. 
Two physical interpretations can be invoked depending on the 
 situation under consideration. 
 Either we consider an impermeable membrane, 
and $p$ would be  
the hydrostatic pressure difference 
between outside and inside; or we choose 
a model of permeable membrane and $p$ would play
the role of an osmotic pressure. Both models 
are equivalent as long as we do not consider 
hydrodynamic flows.

\vspace{0.2cm}
\noindent {\bf (iii) Adhesion forces and boundary terms} \\
The functional derivative induces boundary terms at each contact point. 
The additional variation $\delta E_c^b$ and $\delta E_{w,l}^b$ for the 
curvature, adhesion and tension 
energies, 
 associated with  a small displacement $\delta{\bf r}$ of the 
contact points is given by (see \cite{Cantat99a,Cantatthes})
\begin{eqnarray}
&&\delta E_c^b = \left [\delta \dot{{\bf r}}
\cdot\left ( -{\kappa c \over \sqrt{g}} {\bf n}\right ) \right ]^{L^*}_0 +\left 
[\delta {\bf r}\cdot\left (
\kappa  \drp{c}{s}{\bf n} -{\kappa c^2 \over 2}{\bf t}\right ) \right ]^{L^*}_0 \; , 
\label{courbord} \\
&&\delta E_{w,l}^b=\left [\rule{0cm}{0.4cm}\delta {\bf r}\cdot\left 
(\rule{0cm}{0.4cm}
\zeta {\bf t} + (\zeta - w(x)) \vv{x}\right ) \right ]^{L^*}_0 \; .
\label{adhbord}
\end{eqnarray}
Following the definition of these boundary points,  they remain
on the substrate.
Thus,
the accessible values for $\delta{\bf r}$ is then reduced to $\delta{\bf r} 
\propto {\bf \hat{x}}$. 
Additionally, in order to keep the curvature energy finite,  we impose a 
vanishing 
value for the contact angle $\phi$ between 
the membrane and the substrate (see Fig.3). Within our formulation, 
this constraint 
does not follow  from the energy minimization and  has thus to be added into 
the physical model. More precisely, at the discontinuity point (say $x_2$)
one has to add to the Helfrich energy a term of the form 
$\kappa (\Delta\psi/\Delta s)^2$ which informs
us on how would the vesicle on the adhered part feels, so to
speak, the behavior of the vesicle at the junction
point on the right side. Across the contact point of a vanishing
extent, $\Delta s\rightarrow 0$, while   the angle,  if it had
to have another value than zero, would make a jump leading
to an abnormally increasing curvature energy. We must then
impose a vanishing contact angle.
These various conditions (motion along the wall and a
 vanishing contact angle)  lead to
${\bf n}=-{\bf \hat{y}}$, ${\bf t}=-{\bf \hat{x}}$ and 
$\delta \dot{{\bf r}} \propto {\bf \hat{x}}$. It follows then that
the term proportional to  $\delta \dot{{\bf r}}$ in eq.(\ref{courbord}) vanishes
automatically. 
The second term becomes  $\kappa/2 \left (c_2^2 \delta x_{2}- c_1^2 \delta 
x_{1}\right )$
with $c_1$ and $c_2$ the curvatures at the left and right contact points. 
These terms are counterbalanced by adhesion and tension terms 
(eq.(\ref{adhbord})) leading to the relation 
\begin{equation}
\frac{\delta E}{\delta x_i}=\mp \left ( \kappa \frac{c_i^2}{2}-w(x_i) \right ) 
{\bf \hat{x}} \; ,
\label{Fbord}
\end{equation}
where the $-$ and $+$  signs refer to the rear and fore 
contact points represented
by the subscript $i=1,2$.
At equilibrium, we recover here the relation  $c=\sqrt{2 w/\kappa}$ 
\cite{Seifert91}.

The energy variation given by eq.(\ref{Fbord}) can not really be identified as a 
physical force. 
It corresponds indeed to a {\it geometrical} point displacement. The "force" 
orientation is here
parallel to the substrate, whereas the real force acting on the contact point, 
considered as a material 
point, is expected to be normal to the substrate.
As we have seen above the curvature forces are indeed normal
when applied to a an adjacent piece of the membrane (see eq. \ref{fc}).
The present "force" has the meaning of how much energy would be involved
in displacing the contact point from one position to another.
That geometrical point is by its very nature sitting on the substrate,
so that the "force" associated with its displacement is naturally tangential.


We find it worthwhile to make a short digression. Suppose that 
the angle is not fixed to zero as we did above. More precisely
suppose that the rigidity is so small or the adhesion
is so  large (see below what does
this mean) then the vesicle will be so tense that it would look
like a droplet  outside some length scale $\ell$ to be determined
below (of course within that scale, which is sufficiently
close to the substrate, the matching must be tangential). If we
do not assume  a value
for the angle (that is no relation between ${\bf n}$ or ${\bf t}$ with
${\bf \hat x}$ and ${\bf \hat y}$), and set $\delta {\bf r} \propto 
{\bf \hat x}$ we find from (\ref{Fbord}) that $c=0$ at the contact
(which means a straight line at the contact) and that the angle
between that line and the substrate obeys
\begin{eqnarray}
w_i&=&\left (1-cos(\phi_i)\right )\zeta  \nonumber\\
\phi_i &\sim &\sqrt{{2 w_i \over \zeta}}  \mbox{ ( for small angles )}
\end{eqnarray}
which is nothing but the Young condition. We have
neglected $\kappa \partial c/\partial s$ in comparison
to $w$. The justification is as follows.    $\kappa \partial c/\partial s\sim
\kappa c_0/\ell$, where $c_0$ is the true contact curvature
given by $\sqrt {2 w/ \kappa}$. The approximation
is legitimate provided that the length scale $\ell \gg \sqrt{\kappa/w}$.
The length $\sqrt{\kappa/w}$ is the radius of curvature at contact.
If the scale of interest is outside that internal region, then
the droplet limit is justified.
It must be emphasized however that 
the  effective contact angle is not an intrinsic property of the adhered
membrane, as for a droplet, but it is linked to other parameters
(rigidity, the vesicle scale--on which depends  $\zeta$--, etc... ).
In particular, 
the tension $\zeta$ is fixed by the reduced volume, which is a global property 
of the 
vesicle : different vesicles of the same phospholipid composition,
but with different sizes, may have 
different 
contact angle on the same substrate.

\subsection{Dynamical equation}\label{dynamics}

An important point which must be emphazised when dealing with dynamics
is the identification of the  dissipation sources.
These are the following: (i) the dissipation in the 
membrane
via molecule rotations (very much like liquid crystals where
dissipation is characterized by the Leslie coefficient), (ii) hydrodynamics
flows inside and outside the vesicle, (iii) friction between the 
monolayers, and (iv) bond breaking and restoring with the substrate.
It is well known that dissipation associated with rotation (internal
dissipation) is negligible in practice\cite{Brochard75}, and
for free vesicles (no substrate) hydrodynamics seems to  be the
most important dissipation. Hydrodynamics induces nonlocal 
interactions\cite{Cantat99}
and this will be dealt with extensively in a forthcoming paper.
Our wish in this paper is  to present a pedestrian model, namely a local one,
which allows for a complete analytical solution that  will
help to identify some key ingredients  in the migration process.
A specific dissipation with the substrate will be introduced later.
For the moment we confine our description to the free 
vesicle case.
The local model to be presented here is similar to the so-called
Rousse model \cite{deGennes} in the community of polymers. Indeed, for
a one dimensional contour in a three dimensional space dynamics
becomes local even in the presence of hydrodynamics\cite{remarkrousse}.

The best way to introduce the dynamical law is to consider a dissipation 
function, proportional 
at each point to the  square of the velocity : 
\begin{equation}
F_d=\frac{\eta}{2}\int |{\bf v}|^2 ds \; .
\label{FD}
\end{equation}
The coefficient $\eta$ is here an effective viscosity and  has the dimension of 
a viscosity per unit length.
Its numerical value is estimated by  $\eta = \eta_{wat}/ R \sim 10^2 kg m^{-2} 
s^{-1}$, 
with $\eta_{wat}$ the water viscosity and $R$ a typical vesicle size.

Neglecting inertial terms, the Euler-Lagrange equations become then 
\begin{equation}
-{\delta E \over \delta {\bf r}} ={\delta F_d \over \delta {\bf v}} 
\hspace{0.5cm} 
\Rightarrow \hspace{0.5cm} \eta {\bf v}= {\bf f} \; . 
\label{dyn}
  \end{equation}
As expected, we find a {\it local} proportionality between the membrane
velocity ${\bf v}$ and the membrane force ${\bf f}$, which is a nonlinear
function of position.
In the present picture where the effective tension $\zeta$ is space-independent
no tangential force appears so that physics will only fix the normal
velocity, while the tangential velocity has no physical
meaning as described below.

\vspace{0.2cm}
\noindent {\bf (i) Normal velocity}\\
The normal membrane force is 
 (see eq. (\ref{fc}, \ref{fl}, \ref{fs})) :
 \begin{equation}
f_n=\kappa \left(\drp{^2c}{s^2}+\frac{c^3}{2} \right)-c \zeta-p \; .
\label{force}
  \end{equation}
From the dynamical law (\ref{dyn}) and the membrane forces expression 
(\ref{force}) we obtain the 
normal velocity as a function of $\psi(s)$ :
\begin{equation}
v_n(s)=f_n=\frac{\kappa}{\eta} \left[\drp{^3 \psi}{s^3}+\frac{1}{2} \left  
(\drp{\psi}{s} \right)^3
-\frac{\zeta}{\kappa}\, \drp{\psi}{s}
 -\frac{p}{\kappa} \right]  \; .
 \label{vnloc}
\end{equation} 
It is convenient to write the dynamical equation in terms of the angle
$\psi$ and not the curvature $c$.  The reason  is that the 
boundary conditions are written naturally as a function of $\psi$ (tangential
matching, $\psi=\pm \pi$, and contact curvature $\partial \psi/\partial s=
\sqrt{2w/\kappa}$).

\vspace{0.2 cm}
\noindent  {\bf (ii) Tangential velocity}\\
There is only  one tangential contribution to the membrane forces, $\partial  
\zeta /\partial s$, which 
is zero with the assumption of a uniform tension (see eq.  \ref{fl}). 
This implies that only the total length is conserved, and not the local one. 
A dilatation of a part of the membrane is then permitted, as long as the 
remaining part of the vesicle 
is contracted in order to keep  the total length unchanged (see Fig.4). 
Within  this approximation, there is no energy variation associated to 
tangential motion, and
therefore no forces. 
In other words we consider the vesicle contour as a mathematical curve, loosing 
the concept
of density : only the shape matters, 
independently of the points distribution on the curve.

We could equivalently assume that only the normal velocity contributes to dissipation.
In that  case the dissipation function (\ref{FD}) would take
the form  $F_{dn}=\eta/2 \,  \int |v_n|^2 ds$
and the equation of motion (\ref{dyn}) becomes $ {\bf f} = \eta v_n {\bf n}$. The tangential force
$\partial\zeta /\partial s$ must then vanish and we get automatically that $\zeta=$constant.
 Thus our assumption
of a global Lagrange multiplier can also be viewed as the result of a   
dissipation  due uniquely  to normal displacements. 


As we have already mentioned, tangential displacements do not 
induce a geometrical change.  If the tangential
velocity has no physical meaning (as is the case with
a constant $\zeta$) its choice
should not affect the physics. We are thus at liberty to choose one
which is convenient (very much like a gauge-field invariant formulation in
electrodynamics).
The choice of a gauge is
interpreted as a reparametrization of the curve. 
As seen below the tangential
velocity is fixed by the normal velocity once the gauge
is specified, but there is naturally no feed back
of that tangential velocity on the normal one (the physical one).
This is the crucial difference between this  non physical tangential velocity and 
a physical one that may arise in the general case (as discussed
in a forthcoming paper).
It must be noted  however, that whether a tangential velocity 
is of physical nature or not,
the knowledge of the  normal
velocity is sufficient to describe vesicle dynamics.
It is thus only via its influence on the normal
velocity that a physical tangential velocity would affect the physics (see
below).
Still in that case we can introduce a second tangential
velocity of geometrical nature that corresponds
to the displacement of the representative points
of the curve and not to the {\it material ones} which are
affected by the physical tangential velocity.

In the present model 
the most  convenient parametrization requires a homogeneous points distribution
along the free 
part of the curve, which is  expressed as
$d/dt \, (s(a)/L^*)=0$. This provides  
the  expression for the  "non physical 
tangential velocity" (see appendix in Ref.\cite{Csahok99a}) : 
\begin{equation}
v_t(s)=v_t(0)-\int_0^{s} c \, v_n \, ds + {s \over L^*}
\left( \int_0^{L^*} c \, v_n \, ds + v_t(L^*)-v_t(0) \right) \; .
\label{vt}
\end{equation}
If the free length $L^*$ remains constant during the motion, as happens for 
a stationary regime, eq.(\ref{vt}) fixing the gauge imposes nothing 
but 
a constant distance between two consecutive points on the vesicle. 
The local length conservation (\ref{divv}), which
is physical, seems then to be implied by a gauge!. In reality, once
we have adopted a contant tension --implying  a vanishing tangential physical
velocity--, any point distribution is of purely geometrical nature,
and we could impose another gauge than the above one, without affecting
the physics; the above tangential velocity does not act
on the normal velocity. Had we  considereed $\zeta$ to be non contant,
we would then have obtained a physical tangential velocity, which
would act on the normal one; use of (\ref{divv}) fixes $\zeta (s)$ 
which in turn acts on the value of $v_n$, and then on physics.
In the simplistic model we adopt,
the tangential velocity  is  determined
{\it a posteriori}, independently of the normal velocity. That is why
it is only a non physical reparametrization, a "gauge". 
A remark is in order :
in this situation the question of a rolling or sliding motion does
not make   sense, 
since both motions differ only by a tangential velocity. 

The membrane velocity is given  as a function of $\psi(a)$, $a$ being the 
auxiliary parametrization of the free part of the vesicle,
running from one contact point to the other. 
In order to obtain a closed system we need a relation between the evolution of 
$\psi(a)$ and the velocities. 
The temporal derivative of $\psi$, for a given $a$, is presented in the appendix 
of Ref.\cite{Csahok99a}
\begin{equation}
\left . {\partial \psi \over\partial t } \right ) _a= c\, v_t -\drp{v_n}{s}\; .
\label{dpsi}
\end{equation}

The last step is the determination of the boundary conditions at the
contact with the substrate.

\vspace{0.2 cm}
 {\bf (iii) Contact points velocity} \\
 The motion of the contact point is 
governed by a binding/unbinding 
mechanism, implying
a dissipation law that differs from the bulk dissipation. 
The most natural way for introducing a dissipation law is
the following (with $\Gamma$ a phenomenological dissipation
coefficient)
\begin{equation}
\Gamma\frac{dx_i}{dt}=-\frac{\delta E}{\delta x_i}\; .
\end{equation}
Using the energy variation (\ref{Fbord}) we get the following dynamical law, 
with 
$w_i=w(x_i)$ and $v_i=dx_i/dt$
\begin{equation}
c_1=\sqrt{\frac{2 w_1 + \Gamma v_1}{\kappa}}  \; , \; c_2=\sqrt{\frac{2 w_2 - 
\Gamma v_2}{\kappa}}\; ,
\label{cheq}
  \end{equation}
which constitute the dynamical boundary conditions. 
These out of equilibrium values for the curvature are quite intuitive : 
the unbinding  delay at the rear point induces a larger curvature than at 
equilibrium, 
whereas the binding delay at the fore point induces a smaller curvature.

\section{Transient behavior}

The formalism presented in the first part lends itself very well 
to analytical computation and stationary shape determination, as will appear 
in the following paragraphs. 
Nevertheless, having access to the transient process is highly desirable. 
In particularly, it checks the dynamical stability of an eventual stationary 
behavior, 
obtained after a relaxation. 
The successive vesicle profiles are determined by a direct numerical 
 implementation of the dynamical equations (\ref{vnloc}), (\ref{vt}) and 
(\ref{dpsi}). Unfortunately, numerical instabilities are difficult to avoid
 around each contact point (due to a contact
adhesion potential). A smoother model, without discontinuities, 
is more convenient for such
 an approach. For this reason, in this paragraph devoted to transient processes, 
the adhesion potential 
 will be supposed to be of small, but non vanishing,
range.
We rapidly summarize below the small technical changes arising from this model 
modification.
The chosen  potential profile  is 
\begin{equation}
w({\bf r}) = w_0 (1 + u_0 x)(\frac{y_0^4}{y^4}-\frac{2y_0^2}{y^2})\; ,
\label{ad-pot}
\end{equation}
with the new length $y_0$ fixing the characteristic distance between 
the substrate and the membrane, and $\hat w(x)=-w_0 (1 + u_0 x)$ the minimum of 
the potential interaction, occurring for
$y=y_0$.  It plays the role of the previous $w(x)$. It depends linearly on $x$ 
with an adhesion gradient $u_0$.
The distance $y_0$ is chosen of the order of $10 n m$, which is small enough in 
regard of the vesicle size
to introduce only small variations between both models. 
In this case the parametrization is performed on the whole closed curve, and the 
boundary terms (eqs. \ref{cheq})  are no more relevant. 
The adhesion forces ${\bf f}_w$ are obtained by functional derivation of $\int 
w({\bf r}) ds$ 
leading to 
\begin{equation}
{\bf f}_w = -(c w + \nabla w \cdot {\bf n} ) {\bf n} \; .
\label{dynlaw2}
\end{equation}

Additionally the gauge condition fixing the tangential velocity eq. (\ref{vt}) 
is simplified :
the velocity  $v_t(0)$ is supposed to be zero, without loss of generality, so 
the first term  
 disappears ; 
the last term is proportional to the length variation of the total parametrized 
curve, which is 
zero because we consider the complete profile and no more the free part of the 
curve. 
Thus we obtain for the gauge, replacing eq. (\ref{vt})  :
\begin{equation}
v_t = - \int_0^s ds v_n c \; .
\label{vt2}
\end{equation}
Using equations (\ref{dynlaw2}) and  (\ref{vt2}) we finally get the dynamical 
equation for ${\bf r}$
\begin{equation}
\frac{\partial {\bf r}(a,t)}{\partial t} =
\left[ \kappa \left(
\frac{d^2 c}{ds^2} + \frac{1}{2} c^3 \right) -c w - ({\bf \nabla} w \cdot {\bf 
n})
- p - \zeta c \right] {\bf n}+ v_t {\bf t} \; .
\label{drdt2}
\end{equation}
The Lagrangian multipliers are determined from the following constraint
equations :
\begin{eqnarray}
&&{dL \over dt} = \int c v_n ds =0 \; , \\
&&{dS \over dt} = \int  v_n ds =0  \; .
\label{constr}
\end{eqnarray}
The normal component of the velocity in eq. (\ref{drdt2}) will be denoted by 
convention as $v_n=v_n^0 - p - c \zeta$.
With this notation the equation (\ref{constr}) appears as 
a very simple linear equation system in $\zeta$ and $p$. Its solution provides 
the pressure and tension values :
\begin{eqnarray}
\zeta = \frac{\langle c v_n^0 \rangle -\langle c \rangle \langle v_n^0 \rangle}
{\langle c^2 \rangle - \langle c \rangle^2 } \; , \\
p= -\zeta \langle c \rangle + \langle v_n^0 \rangle \; .
\label{zeta-p}
\end{eqnarray}
with the average defined by
\begin{equation}
\langle \cdots \rangle \equiv \frac{1}{L} \int_0^L ds \cdots.
\label{av}
\end{equation}

We are now in a position to deal with 
the numerical anlysis.  The dynamics is overdamped and 
for this reason {\it local in time}. Starting from   
an arbitrary profile, forward time integration
provides us with the vesicle
evolution. We have checked that (i) for  a free vesicle (no substrate)
the shape (with no external force) tends
towards that obtained by direct energy minimization, (ii) we have
also checked that for a homogeneous substrate an arbitrary initial
shape evolves after some time to the shapes obtained in \cite{Seifert91}
by direct 
energy 
minimization. 

Let us now turn to the non-equilibrium situation ensured by an adhesion
gradient.  Starting from an initial shape, the vesicle 
acquires a non-symmetric shape and  moves in the gradient
direction. After transients have decayed the vesicle acquires
a permanent regime with a constant velocity. Figure 5
shows the shape evolution.


\section{Stationary motion : direct numerical solution} 

The formulation of our problem in terms of a direct stationary
problem is very convenient  both for a systematic study of the 
velocity evolution  as a function of various parameters.
It will also allow us to present a simple analytical solution.
It is convenient here to come back to the 
contact potential model. 
For a vesicle which has attained a stationary shape and velocity $V$
the equations become steady with $V$ as an unknown
parameter.

 For a  stationary motion along the x-axis, normal and tangential velocities can 
be written as
functions of the angle $\psi$ and of the translational velocity $V$ :
\begin{eqnarray}
&&v_n=V \vv{x}\cdot\vv{n}= V \sin \psi \; , \label{vndeV} \\
&&v_t=V \vv{x}\cdot\vv{t}= V \cos \psi \; .
\end{eqnarray}
The shape and velocity are entirely determined from the relation between normal 
velocities 
and forces. 
The equation of motion is  obtained from eqs. (\ref{vnloc},\ref{vndeV}):
\begin{equation}
V \sin \psi =\frac{\kappa}{\eta} \left[\drp{^3 \psi}{s^3}+\frac{1}{2} \left  
(\drp{\psi}{s} \right)^3
-\frac{\zeta}{\kappa}\,\drp{\psi}{s}
 -\frac{p}{\kappa} \right] \; . 
\label{eqmv}
  \end{equation}
  
Let us present briefly a counting argument showing  that the problem   
is well defined.
  Equation (\ref{eqmv}) is a 
nonlinear third order differential equation for 
$\psi$, 
  with 3  parameters to be determined : $\zeta$, $p$ and $V$. So we need 6 
"informations".
  
We have the following equations at our disposal :\\
$\bullet$ 2  geometrical constraints (eqs. (\ref{cos}) and (\ref{sin})) 
:
\begin{equation}
\int_0^{L^*} \cos \psi(s)\, ds = L_{adh}\, , \, 
 \int_0^{L*} - \sin \psi(s) \, ds = 0 
 \label{sincosbis}
\end{equation} 
$\bullet$ 4 boundary equations corresponding
to  the contact angles and their first derivatives
(the dynamical contact curvatures $c_1$ and $c_2$ given by equation 
(\ref{cheq})) 
\begin{equation}
\psi_1\equiv \psi(s=0)=-\pi \, , \hspace{0.3cm} \psi_2\equiv \psi(s=L^*)=\pi \, 
, \hspace{0.3cm}
\left . \drp{\psi}{s} \right)_{s=0}= c_1 \, , \hspace{0.3cm}
 \left .\drp{\psi}{s}\right)_{s=L^*}=c_2
\label{limites}
\end{equation}  
$\bullet$ 1 equation ensuring that 
the  enclosed  surface
is equal to the prescribed area. \\
$\bullet$ 1 equation ensuring that the total length of the curve 
is precisely the prescribed one, $L$, which is related to the two other
lengths  by
\begin{equation}
L=L^*+L_{adh} \; . 
\label{consL}
\end{equation}

\vspace{0.3cm}

There are thus 8 conditions, for only 6 informations needed.
The system seems then to be overdetermined. This is not the case. Indeed
it must be noted that the problem involves additional 
unknowns which are $L^*$ and $L_{adh}$. So in reality 
we have 8 unknown parameters as well. The problem is thus well defined.
 
Once the shape is determined we must in principle evaluate the area
and change the parameter $p$ until the area coincides with 
the prescribed one. But since the area is a conjugate variable
to  $p$ we can fix $p$ --which is more convenient--
and this will fix some area that is treated as free (not imposed
in advance). 
Additionally we are at liberty to prescribe $L$ 
(that fixes some length scale). 
$L_{adh}$ can then be determined if  $L^*$ is known; 
$L_{adh}$ can thus  be removed from the problem 
upon using  eq.(\ref{consL}). The first  constraint (\ref{sincosbis}) 
becomes then  
\begin{equation}
\int_0^{L^*} \cos \psi(s)\, ds = L-L^* \; .
\label{coster}
\end{equation}
In other words prescribing the total length to  $L$ and the pressure
to  $p$ lowers the number of unknowns by two. This is so because
we do not want to have a specific  area, and that $L^*$ and $L_{adh}$
are not independent if we treat the  total length as known.
That is to say we have 
finally 6 fixed boundary conditions or constraints (\ref{sincosbis}-\ref{limites})
and six parameters  which are 
$L^*$, $V$ and  $\zeta$, plus three constants of integration
due to the third order differential equation (\ref{eqmv}).

A convenient way to solve a differential equation of order $n$ is to transform it 
into a set of $n$ first order coupled differential equations. For that purpose  we  set
  $f_1=\psi$,  $\dot{f_1}=f_2$ and  $\dot{f_2}=f_3$ 
where the dot stands for $\partial / \partial s$.
Equation (\ref{eqmv}) then provides us with  the expression for  $\dot{f_3}$  
$\equiv \partial ^3 \psi /\partial s^3$
as a function of $f_1$, $f_2$  :
\begin{equation}
\dot{f_3} =\eta/\kappa \, P_2 \sin f_1 -f_2^3/2 -p/ \kappa + f_2 P_3 / \kappa 
\equiv F \; .
 \end{equation} 
 In order to make visible the quantities which are treated as unknown parameters
 we shall use the symbols $P_i$ (with $i=1,2...$). As stated above
 there are 
three parameters $L^*=P_1$, $V=P_2$ 
and $\zeta=P_3$.
Solution of  a set of  three equations  involves three integration factors.
This means that we have 6 unknowns, as argued in the last paragraph.
Four physical conditions
at the two  end  points (see eq. (\ref{limites})) are known. Two constraints
are imposed 
(eq. (\ref{sincosbis})), and this makes the problem well posed.
Note that conditions (\ref{sincosbis})  have
an integral form. We find it convenient
to rewrite them in  a differential form.
It is easy to realize that by setting
\begin{equation}
f_4(s)\equiv \int_0^s \sin \psi(s')ds'\; ,\; f_5(s)\equiv \int_0^s \cos 
\psi(s')ds'\; ,
\label{f45}
\end{equation} 
we can write
\begin{equation}
\dot{f_4}= \sin \psi \; =\sin f_1 \; ,\; \dot{f_5}= \cos \psi \; =\cos f_1 
\; .
\label{df45}
\end{equation}
These two functions obviously obey
$f_4(0)=0$, $f_5(0)=0$ , whereas at the second boundary we must  impose
$f_4(L^*)=0$ and $f_5(L^*)=L-L^*$ in order to fulfill  the two constraints 
(\ref{sincosbis}, \ref{coster}). This  trick is performed
at the expense of two additional functions $f_4$ and $f_5$ (whose
determinations involve two integration constants). We have thus augmented our system by 2
differential equations of first order. The two additional integration constants are
precisely fixed by the demand $f_4(0)=0$, $f_5(0)=0$, whereas the conditions
$f_4(L^*)=0$ and $f_5(L^*)=L-L^*$ are
substituted to (\ref{sincosbis}).  Finally
the shooting NAG code used here requires to invoke the boundary conditions
for each function $f_i$, with $i\le 5$. The boundary conditions for  each function is invoked
above, except for $f_3$ which represents the second derivative of $\psi$. This
quantity is not known at the boundaries and there is no constraint to be
imposed on it. Let $P_4$ and $P_5$ denote the values of $f_3$ at the two end
points.
We can thus invoke the boundary conditions of $f_3$ and the boundary values
are quantities which are to be determined. That is to say we introduce two
conditions with two additional unknown parameters. We have then in total
ten unknowns and ten conditions. Cast into this form our formulation can
straightforwardly be implemented into a NAG code (code D02HBF).

In summary the problem to be solved can be written in a standard boundary value
problem with unknown parameters :
\begin{equation}
\begin{array}{lll} 
\dot{f_1} = f_2 \; ,\; &f_1(0)= -\pi \; ,\;& f_1(P_1)= \pi \\
\dot{f_2} = f_3  \; ,\; &f_2(0)=c_1  \; ,\; &f_2(P_1)=c_2 \\
\dot{f_3} = F \; ,\;  &f_3(0)= P_4  \; ,\; &f_3(P_1)= P_5\\
\dot{f_4} = \sin f_1 \; ,\; &f_4(0)= 0 \; ,\;& f_4(P_1)= 0\\
\dot{f_5} = \cos f_1\; ,\;& f_5(0)= 0 \; ,\; &f_5(P_1)= L-P_1
\end{array}
\label{systeme}
\end{equation}

Once the problem is solved the vesicle shape is obtained 
by making use of
equations (\ref{dxds}).


If  $w_1=w_2$, the vesicle is at equilibrium on a homogeneous  substrate and 
one obviously expects a vanishing velocity. This comes out automatically
from the above formulation. If we were interested from the beginning 
in an equilibrium problem, we would then not have introduced $V$ as
an unknown parameter. In that case because the profile is symmetric the second
condition (\ref{sincosbis}) is automatically satisfied, and we would then be left with with
nine conditions for nine unknowns.

When   $w_1 \neq w_2$, there is no equilibrium solution for the 
vesicle, which  has to move towards the stronger adhesion region.
If we impose a vanishing velocity there is no way to fulfill
the second condition (\ref{sincosbis}) (a typical profile would
be the one shown on Fig.7) where starting from one end we arrive
at the other end at a different height. Arriving at the same height
can be achieved only for a specific velocity (or at most
a discrete set of solutions), the one we are seeking. Thus the second condition
of eq.(\ref{sincosbis}) (which is parametrized by the set of $P_i$) can
be viewed as 'quantization' condition. This  is  a nonlinear eigenvalue
problem of Barenblat-Zeldovitch type.

The numerical solution reveals an out-of-equilibrium shape which 
is significantly different from 
the equilibrium one, as shown 
on  Fig.6.  We note that the curvature in front of the vesicle
is higher
than the one behind. The reason is that the adhesion energy
is higher in the front part, so  that the curvature/adhesion balance allows
a higher curvature (the vesicle looses curvature energy at the expense
of a stronger adhesion).

\section{Stationary motion : an analytical solution}

  
The advantage offered by the simplistic
picture of our model is the  possibility to  provide analytical results
and thus to shed light on the physical processes that are involved
in the problem of vesicle propulsion.
It turns out that the equation of motion (\ref{eqmv}), if multiplied by 
$\partial^2\psi/\partial s^2$, 
possesses practically a first integral      
\begin{equation}
V\int_0^{L^*} \drp{^2\psi}{s^2} \sin(\psi) ds = {\kappa \over 
\eta} 
\left( [(\partial c/ \partial s )^2/2]^1_2 +{1  \over 2} 
[\,c^4/4\,]^1_2-\frac{\zeta}{\kappa}
[\,c^2/2\,]^1_2-\frac{p}{\kappa}[\,c\,]^1_2 \right) \; .
\end{equation}
Each r.h.s. term has an explicit form as a function of the contact 
curvatures (which  are 
known), and of their
first derivatives, which  have only negligible contribution for swelled vesicle.
The l.h.s. term can be evaluated for a vesicle shape close to a circle. 
The calculation is sent into appendix  and leads to 
\begin{equation}  
\int \drp{^2\psi}{s^2} sin(\psi) ds =4 \pi^2 \frac{L_{adh}}{L^2} \; .
\label{denom}
\end{equation}
Using the dynamical values for the contact curvature  given by eq.(\ref{cheq}),
we  obtain
an explicit expression for the velocity 
\begin{equation} 
V = \frac{L^2 \kappa }{(2\pi)^2\eta L_{adh}}
\left({1  \over 2} [\,c^4/4\,]^1_2-\frac{\zeta}{\kappa}
[\,c^2/2\,]^1_2-\frac{p}{\kappa}[\,c\,]^1_2 \right)\; .
\label{velo}
\end{equation}
In the simple case where  $\Gamma =0 $ (no dissipation
associated with the substrate), expression (\ref{velo}) becomes 
explicit 
and provides a good agreement with numerical solution (see Fig.8).
The analytical expression for the velocity involves only known parameters, 
except 
$L_{adh}$ and
 $\zeta$. For the comparison between numerical and analytical results, we 
took their
 numerical values.

\vspace{0.2cm}
{\bf Limit ${\bf \delta w \ll 1}$}
\vspace{0.2cm}

Another interesting limiting case concerns the small adhesion difference. 
Expansion  of the numerator in eq.(\ref{velo})  to leading order in $\delta w$
yields
\begin{equation}
V \simeq {\delta w\over \eta/A+\Gamma}\;\;\;\;
A\equiv {w\over \kappa }
{R^2 \over L_{adh}} \left [1- { p\over w} \sqrt{\kappa \over 2 w}-{\zeta \over w} \right]
\label{anal}
\end{equation}
where $R=L/2 \pi$.
The influence of the two dissipation coefficients appears then clearly. 
It depends on the quantity $A$, proportional to the ratio $R/L_{adh}$.
The bulk dissipation increases with the adhesion length, which seems to be a 
very 
robust result, as encountered in the model including hydrodynamics 
dissipation \cite{Cantat99b}.
The local dissipation represented
by $\Gamma$ does not depend on the adhesion length. 
Only the two contact points
matter. Note also that the effective dissipation is $\eta/A+\Gamma$.
The bulk dissipation $\eta/A$ and the contact one $\Gamma$ play
a role of resistances (in an electric analogy) which
would be mounted in series.

\section{Discussion and conclusion}
This paper has given a first extensive presentation of the problem
of vesicle migration in haptotaxis. We have 
reduced as much as possible the complexity of the problem in order
to gain some analytical approximate results.
For that purpose we have neglected hydrodynamics which 
induce nonlocal interactions, and adopted a local model
of the Rousse type. The full dynamical problem
has been solved by adopting a powerful gauge-field
invariant formulation. The dynamical code could
account for the transient and the evolution towards
a steady-state solution. In that context an introduction
of an adhesion potential with a finite, albeit small, range
has proven to be necessary in order to circumvent 
numerical instabilities related to the motion 
of the contact point. This code has the advantage
of dealing with various problems not leading necessarily to
permanent motions. For a   stationary situation we could cast the problem
into a standard boundary value one where the migration velocity
appeared as  an  eigenvalue. This problem is akin
to the nonlinear eigenvalue problem of  Barenblat-Zeldovitch type.
A counting argument showed us that the velocity should belong
to a discrete set, only one of them has been identified; we 
speculate that the solution is unique. The problem could
be systematically solved in a fully intrinsic representation
of the contour. For a rather tense vesicle we have provided
an analytical solution which is in a good agreement with the numerical
one.   We have identified the role played by the adhesion length
in selecting the magnitude of the migration velocity even
if no dissipation with the substrate is included. We have
also shown that the bond breaking/restoring dissipation
and the (effective) bulk one are additive  in  a way  which is
analogous to the problem of 
electrical resistances in series. Bulk
dissipation dominates when  
the ratio of 
the bulk dissipation coefficient
to 
the contact one exceeds a certain 
limit, 
which
depends in an intricate manner on various
parameters. 
For real situations, vesicles, and cells in general, are
suspended in aqueous solutions. It is therefore highly important
to include hydrodynamics. Moreover the Lagrange multiplier
$\zeta$ is a local quantity.  We have recently given a brief
account on these questions\cite{Cantat99b}. An extensive discussion
will be presented in the near future.

\appendix

\section{Derivation of equation 44}  

$$
D=\int_0^{L^*} \psi'' \sin \psi ds = -\int_0^{L^*} (\psi')^2 \cos\psi ds$$
We write $\psi = 2\pi s/L^* - \pi + \epsilon$, which implies to  first order 
in 
$\epsilon '$
\begin{eqnarray}
D&=&-\left ({2 \pi \over L^*}\right)^2 \int_0^{L^*}\cos\psi ds - {4 \pi \over 
L^*}
 \int_0^{L^*}\epsilon ' \cos (2\pi s/L^* - \pi + \epsilon) ds \nonumber \\
 &=&-\left ({2 \pi \over L^*}\right)^2 L_{adh}+ {4 \pi \over L^*}
\int_0^{L^*}\epsilon ' \cos (2\pi s/L^* + \epsilon) ds \nonumber 
\end{eqnarray}
We then make use of the following relation:
$$ {d \over ds}(\sin \psi)= -\left({2 \pi \over L^*}+\epsilon '\right) \cos(2\pi 
s/L^*+ \epsilon)$$
The integral between $0$ and $L^*$ of the l.h.s. trem vanishes.  We then
obtain
$$\int_0^{L^*}\epsilon ' \cos (2\pi s/L^* + \epsilon) ds = {2 \pi \over 
L^*}\int_0^{L^*}\cos \psi ds
= {2 \pi \over L^*} L_{adh}$$
The sought after relation has then the form
$$D= \left ({2 \pi \over L^*}\right)^2 L_{adh}$$


\newpage 
\noindent Fig. 1  Notations used in the text. \\  
Fig. 2 A geometrical explanation of the arc length variation with time. \\ 
Fig. 3 Force equilibrium at the fore contact point in the small rigidity limit. \\
Fig. 4 Translation of a circle obtained with  a purely normal motion. The right part is dilated, 
  whereas the left part is contracted.\\ 
Fig. 5 Successive vesicle profiles. The first one  with open circles  is an arbitrarily 
  chosen initial shape. It relaxes to a permanent shape marked by filled circles on a inhomogeneous 
  substrate. \\
Fig. 6 Out of equilibrium adhering vesicle profiles.  $V$ is measured in units of $100\mu m$
  and $W$ in units of $10^{-4} mJ/m^2$.\\ 
Fig. 7 Geometrical constraint on the curve. \\
Fig. 8 Evolution of the vesicle velocity as a function of the adhesion difference.


\begin{thebibliography}{10}
\bibitem{cytoskeletton}
As a leucocyte starts to crawl, part of its fluid cytoplasm turns rigid in a
  kind of sol-gel transition. The neutrophil extends a flat protrusion (leading
  lamella) that attaches to the underlying substrate, primarily through the
  action of membrane-adhesion proteins. The adhesion with the substrate
  provides a traction force that enables the cell to pull itself forward. The
  molecular motors provide a driving force to the actin filament that pushes
  the cell forward.
\bibitem{remarkrousse}
The reason for locality is the following. For a true surface in 3D any motion
  on some scale $R$ induces a perturbation at least on that scale that couples
  different points on the surface. For 1D object in 3D circulation of the
  liquid can wander around the line (or polymer like geometry) without needing
  to affect distant points, leading thereby to local dynamics.
\end{thebibliography}
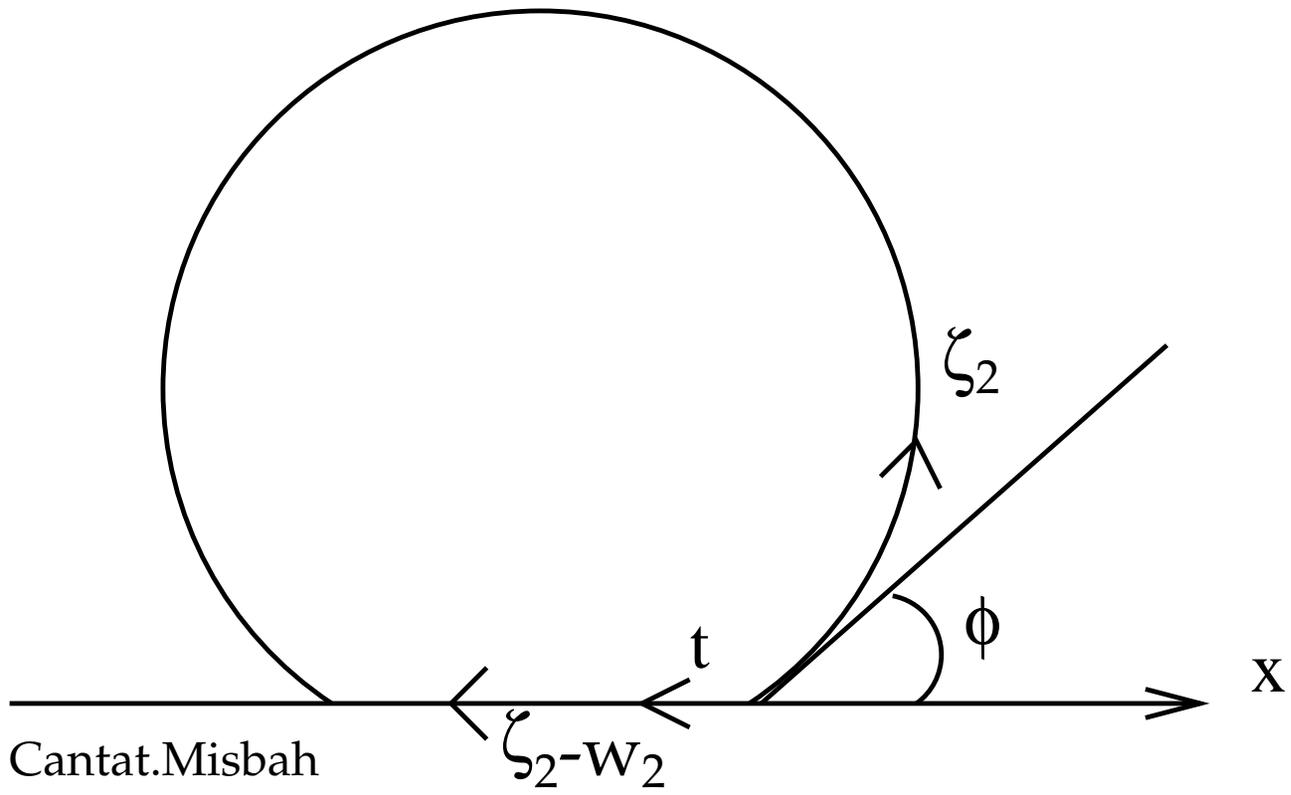
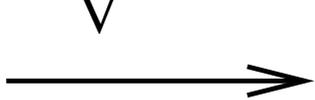
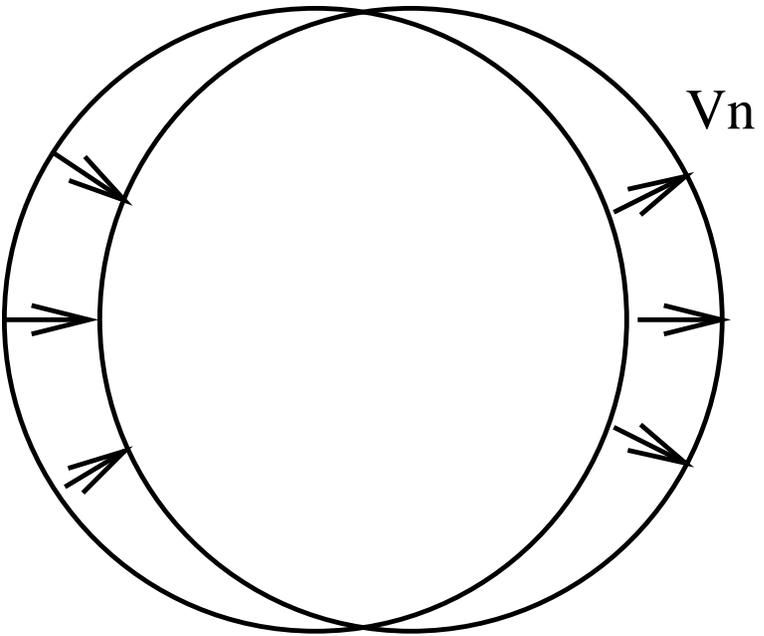
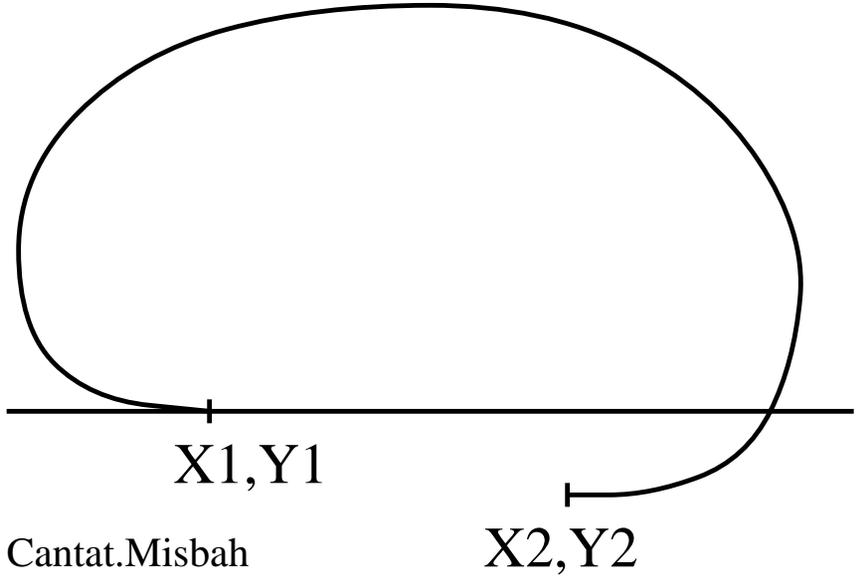
\end{document}